\documentclass[12pt]{article}
\usepackage{amssymb}
\usepackage[dvips]{epsfig}
\usepackage[portuges]{babel}
\usepackage[utf8]{inputenc}
\usepackage{slashed}

\usepackage{color}
\usepackage{comment}
\usepackage{cite}

\newcommand{\be}{\begin{equation}}
\newcommand{\ee}{\end{equation}}
\def\bea{\begin{eqnarray}}
\def\eea{\end{eqnarray}}
\newcommand{\bn}{\begin{eqnarray}}
\newcommand{\en}{\end{eqnarray}}

\usepackage{amsmath}

\DeclareMathOperator{\I}{i}
\DeclareMathOperator{\E}{e}

\usepackage{indentfirst}

\begin{document}
\title{\textbf{Spinning cosmic strings in Brans-Dicke \\ gravity as the generators of the \\ rotational curves of the galaxies}\footnote{Work presented at the 23$^\text{a.}$ Mostra de Ensino, Pesquisa e Extensão, IFRS, Câmpus Porto Alegre, 2023.} \\ \textcolor{white}{a} \\
	\textbf{Cordas cósmicas girantes na gravitação de Brans-Dicke como geradoras das \\ curvas de rotação de galáxias}$^*$}
\author{João Victor Chaves$^{\dag}$, Sérgio Mittmann dos Santos$^{\dag\dag}$ \\
\textit{{\small Instituto Federal de Educação, Ciência e Tecnologia do Rio Grande do Sul -- IFRS}} \\
\textit{{\small Câmpus Porto Alegre}} \\
\small{$^{\dag}$\texttt{victorchhaves@gmail.com,} $^{\dag\dag}$\texttt{sergio.santos@poa.ifrs.edu.br}}
\\}
%\date{\today}
\date{}
\maketitle

%\begin{abstract}
\textbf{Abstract} Only 5\% of what makes up the Universe is well understood, and it consists of baryonic matter and radiation. \textit{Dark matter and energy} correspond to the remaining 95\% of the Universe, and their origin and evolution have not yet been satisfactorily explained. Dark matter, supposedly present in the galaxies' halo region, appears to be the mechanism that causes the unusual behavior of the stars' tangential velocity, which is higher than that predicted by the interaction with visible matter. With the solutions of the equations of motion for spacetime generated by a spinning cosmic string with an internal structure in Brans-Dicke gravitation, the present work aimed to evaluate whether this type of string can play the role currently defined as that of dark matter, which is to originate the typical rotational curves of galaxies, responsible for maintaining the tangential velocities of the stars that form these galaxies, whose behavior cannot be justified solely by the observed baryonic matter. For this, the model was used to obtain the velocities of the stars of 4 Sc-type galaxies, to be compared with their respective experimentally observed values.
\\ \\
{\normalsize \textbf{Keywords} \textit{Spinning cosmic strings, Brans-Dicke gravitation, Rotational curves of galaxies}}
%\end{abstract}
\\ \\
%\begin{abstract}
\textcolor{white}{mai}\textbf{Resumo} Apenas 5\% do que compõe o Universo é bem compreendido, e consiste em matéria bariônica e radiação. \textit{Matéria e energia escuras} correspondem aos demais 95\% do Universo, e ainda não têm as suas origem e evolução satisfatoriamente explicadas. A matéria escura, supostamente presente na região do halo das galáxias, parece ser o mecanismo que causa o comportamento incomum da velocidade tangencial das estrelas, que é superior ao previsto pela interação com a matéria visível. Com as soluções das equações de movimento para o espaço-tempo gerado por uma corda cósmica girante com estrutura interna na gravitação de Brans-Dicke, o presente trabalho objetivou avaliar se esse tipo de corda pode desempenhar o papel atualmente definido como sendo da matéria escura, qual seja o de originar as típicas curvas de rotação das galáxias, responsáveis pela manutenção das velocidades tangenciais das estrelas que formam essas galáxias, cujo comportamento não pode ser justificado apenas pela matéria bariônica observada. Para isso, o modelo foi utilizado para a obtenção das velocidades das estrelas de 4 galáxias do tipo Sc, para serem comparadas com os seus respectivos valores observados experimentalmente.
\\ \\
{\normalsize \textbf{Palavras-chaves} \textit{Cordas cósmicas girantes, Gravitação de Brans-Dicke, Curvas de rotação de galáxias}}
%\end{abstract}

%{\small Palavras-chave \textit{Sistemas de equações diferenciais acopladas, Equações de mo\-vi\-men\-to, Gra\-vi\-ta\-ção de Brans-Dicke, Curvas tipo-tempo fechadas}}

\section{Introdução}
\label{Introdução}

Atualmente, há a compreensão de apenas 5\% do que compõe o Universo, consistindo de matéria bariônica e radiação. \textit{Matéria e energia escuras}, que correspondem aos demais 95\% do Universo, ainda não têm as suas origem e evolução satisfatoriamente explicadas. A matéria escura, supostamente presente na região do halo das galáxias, parece ser o mecanismo que causa o comportamento incomum da velocidade tangencial das estrelas, que é superior ao previsto pela interação com a matéria visível \cite{Zwicky1933, Zwicky1937, Rubin1980, Persic1995, Salucci1997}. Além dessa velocidade maior do que a esperada, à medida que a distância das estrelas a partir do centro da galáxia cresce, suas velocidades tendem a um valor constante \cite{Rubin1980}. Uma explicação alternativa a esse comportamento é considerar uma teoria escalar-tensorial que possa resolver o paradigma das chamadas \textit{curvas de rotação} das galáxias. Os resultados das referências \cite{Matos2002, Lee2004, Costa2006} reproduzem o comportamento da velocidade à grande distância, com a ordem de grandeza esperada. No entanto, eles não fornecem informações sobre como a intensidade dessa velocidade evolui. As soluções para o espaço-tempo de uma fonte com simetria associada a uma \textit{corda cósmica girante} na \textit{gravitação de Brans-Dicke} (\textit{BD}) se mostraram adequadas para gerar as curvas de rotação de 21 galáxias do tipo \textit{Sc} (a Via Láctea é uma galáxia do tipo Sbb, e veja, por exemplo, \cite{Oliveira2004}), pois reproduzem o comportamento do crescimento da velocidade tangencial na ordem de grandeza compatível com os respectivos dados experimentais \cite{Santos2018A}. A opção pela gravitação de BD \cite{Brans1961, Dicke1962} no trabalho \cite{Santos2018A} deveu-se ao fato de que a origem das cordas cósmicas foi logo após o Big Bang, quando ainda era importante a presença de um campo escalar
\cite{Damour1993, Damour1993A, Damour1994, Contaldi1999}, como o introduzido por Brans e Dicke, na década de 1960, diferente do que ocorre atualmente, quando a dinâmica de tal campo pode ser desprezada, e a \textit{Relatividade Geral} (\textit{RG}) de Einstein é a teoria gravitacional mais apropriada para a descrição dos fenômenos observados no Universo.

Porque as soluções do trabalho \cite{Santos2018A} não levaram em consideração a estrutura interna de uma corda cósmica girante, não foi possível concluir que esse tipo de corda fosse a geradora do espaço-tempo encontrado. Neste sentido, o objetivo do projeto \textit{Espaços-tempos gerados por cordas cósmicas girantes na gravitação de Brans-Dicke}, vinculado aos editais IFRS 5/2018 e 64/2019, foi o de incluir a estrutura interna das cordas cósmicas girantes nas equações que modelaram o espaço-tempo obtido, a fim de verificar se são elas efetivamente as geradoras desse espaço-tempo. No desenvolvimento desse projeto, foram obtidas as novas equações de movimento, as quais formam um sistema composto por 7 equações diferenciais acopladas
\cite{Chaves2018, Oliveira2021}. Devido à grande complexidade desse sistema das novas equações de movimento, o mesmo somente foi resolvido no projeto \textit{Soluções gerais para as equações de movimento com uma métrica não diagonal na gravitação de Brans-Dicke}, vinculado ao edital IFRS 58/2020, com a introdução de um método alternativo à resolução de sistemas de equações diferenciais acopladas \cite{Chaves2022}. Os resultados alcançados nesses 2 projetos mencionados aqui são mostrados nas Seções \ref{Definição} e \ref{A11}.

Agora, com as soluções das equações de movimento para o espaço-tempo gerado por uma corda cósmica girante (com estrutura interna) obtidas no trabalho \cite{Chaves2022}, o trabalho apresentado aqui objetivou avaliar se esse tipo de corda pode de fato desempenhar o papel atualmente definido como sendo da matéria escura, que é o de originar as típicas curvas de rotação das galáxias, responsáveis pela manutenção das velocidades tangenciais das estrelas que formam essas galáxias, cujo comportamento não pode ser justificado apenas pela matéria bariônica presente. 
A Seção \ref{vvarphi} apresenta o teste do modelo aplicado nas 
4 galáxias do tipo Sc destacadas no artigo \cite{Santos2017}, o qual apresenta as principais ideias do trabalho \cite{Santos2018A}.
Na Seção \ref{Considerações}, são expostas as considerações finais.

\section{Definição das equações de movimento e nova proposta para as suas soluções}
\label{Definição}

No referencial de Einstein, as equações de movimento da gravitação de BD são escritas como
\begin{eqnarray}
	G_{\mu\nu}&=& 2 \partial_\mu \phi \partial_\nu \phi - g_{\mu\nu} g^{\rho \sigma} \partial_{\rho} \phi \partial_\sigma \phi + \frac{8\pi}{\phi}\; T_{\mu\nu} \; , \label{eqs_movimento_BD_Einstein_1}
	\\
	\Box \phi &=&
	%\frac 1 {\sqrt{-g}} \partial_\mu \left ( \sqrt{-g} \partial^\mu \phi \right ) =
	\frac {8\pi}{2\omega +3}\; T \; .
	\label{eqs_movimento_BD_Einstein_2}
\end{eqnarray}
O número de termos não nulos da métrica $g_{\mu\nu}$ estabelece quantas equações de movimento a equação (\ref{eqs_movimento_BD_Einstein_1}) dará origem.
Assim, o sistema das equações de movimento é formado pelas equações originadas a partir da equação (\ref{eqs_movimento_BD_Einstein_1}), junto com a equação (\ref{eqs_movimento_BD_Einstein_2}).
O campo $\phi$ visto aqui é em função da coordenada radial $r$, ou seja, $\phi= \phi(r)$, e, se não há massa no espaço-tempo gerado pela estrutura estudada, o \textit{tensor energia-momento} $T=0$.
\begin{comment}
Quando o campo $\phi$ é em função da coordenada radial $r$, ou seja, $\phi= \phi(r)$, e o \textit{tensor energia-momento} $T=0$, isto é, se não há massa no espaço-tempo gerado pela estrutura estudada, a equação (\ref{eqs_movimento_BD_Einstein_2}) pode ser escrita como
\begin{equation}
\phi '+ r \, \phi'' = 0 \; , \label{eqs_movimento_BD_Einstein_2_Tzero}
\end{equation}
onde as derivadas são também em função de $r$.
A solução geral da equação (\ref{eqs_movimento_BD_Einstein_2_Tzero}) é
\begin{equation}
\phi= C_2 + C_1 \ln r\; ,
\label{solucaogeralphi}
\end{equation}
onde $C_1$ e $C_2$ são constantes de integração.
Essa solução (\ref{solucaogeralphi}) está acoplada às equações diferenciais obtidas a partir da equação (\ref{eqs_movimento_BD_Einstein_1}).
\end{comment}

Quando $g_{\mu\nu}$ é a métrica geral com simetria cilíndrica \cite{Jensen1992}
\begin{equation}\label{metricagmunu}
	g_{\mu\nu}= \left[
	\begin{array}{c c c c}
		g_{tt} & 0 & 0 & g_{t\varphi} \\
		0 & g_{rr} & 0 & 0 \\
		0 & 0 & g_{zz} & 0 \\
		g_{\varphi t} & 0  & 0 & g_{\varphi\varphi}
	\end{array}
	\right]=
	\left[
	\begin{array}{c c c c}
		-\E^{2\alpha} & 0 & 0 & -\E^\alpha M \\
		0 & \E^{2\left(\beta-\alpha\right)} & 0 & 0 \\
		0 & 0 & \E^{2\left(\beta-\alpha\right)} & 0 \\
		-\E^\alpha M & 0  & 0 & \E^{-2\alpha}r^2-M^2
	\end{array}
	\right] \; ,
\end{equation}
onde $\alpha$, $\beta$ e $M$ são funções da coordenada $r$, as soluções das equações (\ref{eqs_movimento_BD_Einstein_1}) e (\ref{eqs_movimento_BD_Einstein_2}), a partir do sistema formado por 7 equações diferenciais acopladas (6 equações originadas da equação (\ref{eqs_movimento_BD_Einstein_1}), e mais a equação (\ref{eqs_movimento_BD_Einstein_2})), com $T=0$, são \cite{Santos2018}
\begin{eqnarray}
	\beta &=& 2\alpha \; , \label{solucaobeta} \\
	\alpha &=& a_2 \ln r + a_3 \; , \label{solucaoalphagmunu} \\
	M &=& a_1 \E^{a_3} r^{a_2} \; , \label{solucaoMgmunu} \\
	\phi &=&  \phi_0 \pm \sqrt{a_2 \left( 2-a_2\right)} \ln r \; , \label{solucaophigmunu}
\end{eqnarray}
onde $a_1$, $a_2$ e $a_3$ e $\phi_0$ são constantes de integração.
A solução (\ref{solucaobeta}) foi definida para a manutenção da invariância de Lorentz, o que permite ainda reescrever a métrica (\ref{metricagmunu}) como
\begin{equation}\label{metricagmunu2alpha}
	g_{\mu\nu}=
	\left[
	\begin{array}{c c c c}
		-\E^{2\alpha} & 0 & 0 & -\E^\alpha M \\
		0 & \E^{2\alpha} & 0 & 0 \\
		0 & 0 & \E^{2\alpha} & 0 \\
		-\E^\alpha M & 0  & 0 & \E^{-2\alpha}r^2-M^2
	\end{array}
	\right] \; .
\end{equation}
\begin{comment}
Comparando as soluções (\ref{solucaogeralphi}) e (\ref{solucaophigmunu}), conclui-se também que, com a métrica (\ref{metricagmunu}) (ou (\ref{metricagmunu2alpha})),
\begin{eqnarray}
C_1 &=& \pm \sqrt{a_2 \left( 2-a_2\right)} \; , \\
C_2 &=& \phi_0 \; .
\end{eqnarray}
\end{comment}

Aqui, é introduzida a métrica também com simetria cilíndrica, como a métrica (\ref{metricagmunu2alpha}),
\begin{equation}\label{metricagmunuA}
	g_{{\mu\nu}_A}=
	\left[
	\begin{array}{c c c c}
		-A_1 & 0 & 0 & A_3 \\
		0 & A_1 & 0 & 0 \\
		0 & 0 & A_1 & 0 \\
		A_3 & 0  & 0 & A_2
	\end{array}
	\right] \; ,
\end{equation}
onde $A_1$, $A_2$ e $A_3$ são funções quaisquer da coordenada radial $r$, do campo $\phi$, ou ainda funções constantes.
Com essa métrica, como na métrica (\ref{metricagmunu}), são previstas 7 equações de movimento, porque 6 também são os seus termos não nulos.
A escolha $g_{tt}= -g_{zz}= -A_1$ é para que seja igualmente, como na métrica (\ref{metricagmunu2alpha}), preservada a invariância de Lorentz.
Além disso, é considerado ainda $g_{rr}= g_{zz}= A_1$.
As funções $A_n$, com $n=1,2,3$, asseguram uma maior generalidade à métrica (\ref{metricagmunuA}), além daquela apresentada pela
%quando comparada com a
métrica (\ref{metricagmunu}).

Igualando as métricas (\ref{metricagmunu2alpha}) e (\ref{metricagmunuA}), pode ser escrito um sistema de equações para a obtenção das soluções gerais das equações de movimento com as funções $A_n$, isto é, considerando
\begin{equation}
\left[
\begin{array}{c c c c}
	-\E^{2\alpha} & 0 & 0 & -\E^\alpha M \\
	0 & \E^{2\alpha} & 0 & 0 \\
	0 & 0 & \E^{2\alpha} & 0 \\
	-\E^\alpha M & 0  & 0 & \E^{-2\alpha}r^2-M^2
\end{array}
\right] =
\left[
\begin{array}{c c c c}
	-A_1 & 0 & 0 & A_3 \\
	0 & A_1 & 0 & 0 \\
	0 & 0 & A_1 & 0 \\
	A_3 & 0  & 0 & A_2
\end{array}
\right] \; ,
\end{equation}
temos o sistema
\begin{eqnarray}
A_1 &=& \E^{2\alpha_A} \; ,  \\
A_2 &=& \E^{-2\alpha_A}r^2-M_A^{\; 2} \; ,  \\
A_3 &=& -\E^{\alpha_A} M_A \; ,
\end{eqnarray}
onde o índice $A$ inserido nas funções $\alpha$ e $M$ indica que as novas soluções para essas funções serão escritas em função das funções $A_n$, o que as distinguem das soluções (\ref{solucaoalphagmunu}) e (\ref{solucaoMgmunu}).
Além disso, se o campo $\phi$ compõe uma ou mais das funções $A_n$, ainda haverá a solução $\phi_A$, que, diferente da solução (\ref{solucaophigmunu}), depende também dessas funções $A_n$.
Resumindo, as soluções para o sistema formado pelas equações (\ref{eqs_movimento_BD_Einstein_1}) e (\ref{eqs_movimento_BD_Einstein_2}), originado com a métrica (\ref{metricagmunuA}),
%e junto com a solução (\ref{solucaogeralphi}),
podem ser escritas como
\begin{eqnarray}
\alpha_A &=& \alpha_A(A_1, A_2, A_3) \; , \label{solucaoalphaA} \\
M_A &=& M_A(A_1, A_2, A_3) \; , \label{solucaoMA} \\
\phi_A &=& \phi_A(A_1, A_2, A_3) \; . \label{solucaophiA}
\end{eqnarray}

Agora, pode ser escrito o sistema
\begin{eqnarray}
\alpha &=& \alpha_A \; , \label{alphaalphaA} \\
M &=& M_A \; , \label{MMA} \\
\phi &=& \phi_A \; , \label{phiphiA}
\end{eqnarray}
onde o lado esquerdo das equações corresponde às soluções (\ref{solucaoalphagmunu})-(\ref{solucaophigmunu}), e o lado direito às soluções (\ref{solucaoalphaA})-(\ref{solucaophiA}).
Dessa forma, o sistema (\ref{alphaalphaA})-(\ref{phiphiA}) permite determinar $a_1$, $a_2$, $a_3$ e $\phi_0$ que satisfaçam o sistema das soluções (\ref{solucaoalphagmunu})-(\ref{solucaophigmunu}), embora possam não necessariamente continuarem sendo constantes, mas serem funções de $r$, isto é, há a possibilidade de que $a_1=a_1\left( r \right)$, $a_2=a_2\left( r \right)$, $a_3=a_3\left( r \right)$ e $\phi_0=\phi_0\left( r \right)$, porque, agora, carregam as informações da nova métrica $g_{{\mu\nu}_A}$.
O interessante neste procedimento está no fato de que, dependendo obviamente dos termos da métrica $g_{{\mu\nu}_A}$, a obtenção das soluções das equações de movimento torna-se muito mais simples do que a complexa tarefa de resolver um sistema com 7 equações diferenciais acopladas.
Na próxima seção, é exemplificada uma aplicação da proposta aqui mostrada.

\section{Soluções das equações de movimento para
%quando
$A_1= 1$}
\label{A11}

Os trabalhos \cite{Santos2018, Santos2017} exploraram o espaço-tempo formado a partir de uma simetria associada como a que é prevista para as cordas cósmicas girantes, mas não foi considerada a estrutura interna dessas cordas.
Sendo assim, não é possível afirmar que a formação efetiva dos \mbox{espaços-tempos} seja originada a partir das cordas cósmicas girantes.
Neste sentido, o objetivo principal dos trabalhos \cite{Chaves2018, Oliveira2021} foi o de incluir a estrutura interna das cordas cósmicas girantes nas equações que modelaram os espaços-tempos obtidos nos trabalhos \cite{Santos2018, Santos2017}, a fim de verificar se são elas que de fato geram esses espaços-tempos.
Para a construção da métrica com a estrutura interna, partiu-se do elemento de linha %(\ref{eq_3_mazur})
\cite{Mazur1986}
\begin{equation}
	ds^2  =  -\left ( dt+4GJ \; d\varphi \right )^2 + r^2 \alpha^2 \; d\varphi^2 +dr^2 +dz^2 \; ,
	\label{eq_3_mazur}
\end{equation}
onde
\begin{equation}\label{alpha}
	\alpha  =  1-4G\mu_s \; ,
\end{equation}
e a massa $\mu_s$ e o momento angular $J$ são ambos por unidade de comprimento e constantes.
Assim, na gravitação de BD, quando $G\rightarrow \frac{1}{\phi}$, o elemento de linha (\ref{eq_3_mazur}) é reescrito como
\begin{equation}
	ds^2  =  -\left ( dt + \frac{4J}{\phi} \; d\varphi \right )^2 + r^2 \left( 1-\frac{4\mu_s}{\phi} \right)^2 \; d\varphi^2 +dr^2 +dz^2 \; ,
	\label{eq_3_mazur_BD}
\end{equation}
onde o campo escalar $\phi$ é uma função apenas da coordenada radial $r$, isto é, $\phi = \phi\left(r\right)$, e
\begin{equation}\label{metricaMazurBD}
	g_{\mu\nu}=
	\left[
	\begin{array}{c c c c}
		-1 & 0 & 0 & -\;\frac{4J}{\phi} \\
		0 & 1 & 0 & 0 \\
		0 & 0 & 1 & 0 \\
		-\;\frac{4J}{\phi} & 0  & 0 & \;r^2 -\;\frac{8\mu_s r^2}{\phi} + \frac{16}{\phi^2} \left( \mu_s^{\;2} r^2 - J^2 \right)\;
	\end{array}
	\right] \; ,
\end{equation}
com $g_{11}= g_{tt}$, $g_{22}= g_{rr}$, $g_{33}= g_{zz}$, e $g_{44}= g_{\varphi\varphi}$.
Com a métrica (\ref{metricaMazurBD}), foram definidas as equações de movimento para as cordas cósmicas girantes na gravitação de BD, as quais são mostradas no sistema formado pelas equações (\ref{eq_ei_BD_1})-(\ref{eq_ei_BD_7}).
\begin{eqnarray}
	\mathcal{G}^t_{\;\;t}&:& \phi'^{2\;}+f_{tt} =0 \label{eq_ei_BD_1} \; , \\
	\mathcal{G}^t_{\;\;\varphi}&:& f_{t\varphi} =0 \label{eq_ei_BD_2} \; , \\
	\mathcal{G}^r_{\;\;r}&:& \frac{4J^2}{r^2 \phi^2 \left( \phi-4\mu_s \right)^2}-1=0 \label{eq_ei_BD_3} \; , \\
	\mathcal{G}^z_{\;\;z}&:& \phi'^{2\;}+f_{zz} =0 \label{eq_ei_BD_4} \; , \\
	\mathcal{G}^\varphi_{\;\;t}&:& \phi'\left[ \frac{1}{r}+\frac{2\left( \phi-2\mu_s\right)\phi'}{\phi\left( \phi-4\mu_s \right)}\right]-\phi'' = 0 \label{eq_ei_BD_5} \; , \\
	\mathcal{G}^\varphi_{\;\;\varphi}&:& \phi'^{2\;}+f_{\varphi\varphi} =0 \label{eq_ei_BD_6} \; , \\
	\Box \phi &:& \phi'' +\phi' \left[ \frac{1}{r} + \frac{4\mu_s\phi'}{\phi\left( \phi-4\mu_s \right)} \right] = 0 \label{eq_ei_BD_7} \; ,
\end{eqnarray}
onde as derivadas são em função da coordenada $r$, isto é, $\phi'=\frac{d\phi}{dr}$ e $\phi''=\frac{d^2\phi}{dr^2}$,
e $f_{tt}$, $f_{t\varphi}$, $f_{zz}$ e $f_{\varphi\varphi}$ são funções de $\mu_s$, $J$, $r$, $\phi$, $\phi'$ e $\phi''$, ou seja,
\begin{eqnarray}
	f_{tt} &=& f_{tt} \left( \mu_s, J, r, \phi, \phi', \phi'' \right) \label{eq_f_tt} \; , \\
	f_{t\varphi} &=& f_{t\varphi} \left( \mu_s, J, r, \phi, \phi', \phi'' \right) \label{eq_f_tvarphi} \; , \\
	f_{zz} &=& f_{zz} \left( \mu_s, J, r, \phi, \phi', \phi'' \right) \label{eq_f_zz} \; , \\
	f_{\varphi\varphi} &=& f_{\varphi\varphi} \left( \mu_s, J, r, \phi, \phi', \phi'' \right) \label{eq_f_varphivarphi} \; .
\end{eqnarray}
As equações (\ref{eq_ei_BD_1})-(\ref{eq_ei_BD_7}) podem ser vistas completas, inclusive com as funções (\ref{eq_f_tt})-(\ref{eq_f_varphivarphi}) mostradas integralmente, na referência \cite{Santos2021}.
Apesar dos esforços empregados durante o desenvolvimento dos trabalhos \cite{Chaves2018, Oliveira2021}, as soluções gerais do sistema (\ref{eq_ei_BD_1})-(\ref{eq_ei_BD_7}) não foram encontradas, por causa da sua grande complexidade.

Agora, com a proposta descrita na Seção \ref{Definição}, é possível resolver o sistema (\ref{eq_ei_BD_1})-(\ref{eq_ei_BD_7}).
Inicialmente, com a métrica (\ref{metricaMazurBD}), pode ser identificado que as funções $A_n$ são
\begin{eqnarray}
A_1 &=& 1 \; , \label{A1MazurBD} \\
A_2 &=& r^2 -\;\frac{8\mu_s r^2}{\phi} + \frac{16}{\phi^2} \left( \mu_s^{\;2} r^2 - J^2 \right) \; , \\
A_3	&=& -\;\frac{4J}{\phi} \; , \label{A3MazurBD}
\end{eqnarray}
O sistema (\ref{A1MazurBD})-(\ref{A3MazurBD}), considerando $C_1= 0, \pm 1, \pm 2, \pm 3, \,\ldots$, leva às soluções
\begin{align}
1&:& \alpha_A &= 2\pi\I C_1 \; , & M_A &= \frac{2J}{\mu_s}\; , & \phi_A &= 2\mu_s \; ; \\
2&:& \alpha_A &= \pi\I \left( 1+2C_1 \right)\; , & M_A &= -\;\frac{2J}{\mu_s}\; , & \phi_A &= 2\mu_s \; .
\end{align}
Para a solução real de $\alpha_A$, $C_1= 0$, e o sistema (\ref{alphaalphaA})-(\ref{phiphiA}) pode ser escrito como
\begin{eqnarray}
	a_2 \ln r + a_3 &=& 0 \; , \\
	a_1 \E^{a_3} r^{a_2} &=& \;\frac{2J}{\mu_s} \; , \\
	\phi_0 \pm \sqrt{a_2 \left( 2-a_2\right)} \ln r &=& 2\mu_s \; ,
\end{eqnarray}
o que permite determinar que
\begin{eqnarray}
a_1 &=& \;\frac{2J}{\mu_s} \; , \\
a_2 = a_3 &=& 0 \; , \\
\phi_0 &=& 2\mu_s \; .
\end{eqnarray}
Dessa forma, após a métrica (\ref{metricaMazurBD}) ser reescrita sob a forma da métrica (\ref{metricagmunuA}) (introduzida no método proposto na Seção \ref{Definição}), as equações de movimento (\ref{eq_ei_BD_1})-(\ref{eq_ei_BD_7}) são adequadamente satisfeitas quando a solução para o campo $\phi$ é
\begin{equation} \label{solucaophiMazurBD}
\phi= 2\mu_s \; .
\end{equation}
Além disso, é possível verificar que pode haver CTCs no espaço-tempo gerado por cordas cósmicas girantes, quando
\begin{equation}
%r< \sqrt{\frac{4 J^2}{\mu_s^{\; 2}}} \; ,
r
%_{CTC}
< \frac{1}{c}\;\frac{2 J}{\mu_s} \; ,\;\;\;\;\;\;\;\;\; c= v_{luz}= 2\text{,}998\cdot 10^8 \text{ m/s}\; ,
\end{equation}
porque essa é a região onde $g_{\varphi\varphi}< 0$.
\begin{comment}
Como
\begin{equation}
%J= \frac{L}{l}= \frac{I\; \omega}{l}= \frac{1}{2} \;\mu_s r_s^{\; 2}\; \frac{2\pi}{T} \; ,
J= \frac{L}{l}= \frac{I\; \omega}{l}= \frac{\frac{1}{2}\;m\; r_s^{\; 2}\; \omega}{l}= \frac{1}{2} \;\mu_s r_s^{\; 2}\; 2\pi f \; ,
\end{equation}
e com $r=r_{Bohr}= 5\text{,}292\cdot 10^{-11}$ m $\gg r_s= 10^{-31}$ m, $c= 2$,$998\cdot 10^8$ m/s, e $\mu_s= 10^{21}$ kg/m,
%, e $r_s= 10^{-31}$ m,
%e $T=1$ s,
\begin{equation}
f> 2\text{,}525\cdot 10^{59}\text{ Hz}\; , \nonumber
%r_{CTC} < 2\text{,}096\cdot 10^{-70}\text{ m} \ll r_s= 10^{-31}\text{ m}}\; .\nonumber
\end{equation}
o que parece indicar que é improvável que ocorram CTCs em  cordas cósmicas girantes que mantêm a invariância de Lorentz (isto é, quando $\beta=2\alpha$), porque essas cordas, se apresentarem CTCs, são extremamente instáveis.
\end{comment}

\section{Velocidades tangenciais $v_\varphi$ no espaço-tempo gerado por cordas cósmicas girantes na gravitação de \mbox{Brans-Dicke}}
\label{vvarphi}

Com a solução (\ref{solucaophiMazurBD}), o elemento de linha (\ref{eq_3_mazur_BD}) pode ser reescrito como 
\begin{equation}
	ds^2 = -\left( dt+\frac{2J}{\mu_s} \, d\varphi \right)^2+r^2\,  d\varphi^2 + dr^2+dz^2 \; .
	\label{eq_3_mazur_BD_phi2mus}
\end{equation} 
Assim, a densidade lagrangeana de uma partícula no espaço-tempo gerado por uma corda cósmica girante na gravitação de Brans-Dicke, onde a simetria de Lorentz não é quebrada, é 
igual a \cite{Matos2002, Jensen1992}
\begin{equation}
	2 \mathcal{L}=
	-\left( \dot{t}+\frac{2J}{\mu_s} \, \dot{\varphi} \right)^2+r^2\,  \dot{\varphi}^2 + \dot{r}^2+\dot{z}^2 \; ,
	\label{Lagrangiana}
\end{equation}
onde o ponto ($\dot{\color{white}\mu}$) denota a derivada em relação ao \textit{tempo próprio} $\tau$.
Os momentos canônicos associados com essa densidade lagrangeana,
$p_{x^\mu}=\frac {\partial\mathcal{L}}{\partial\dot{x}^\mu}$,
são
\begin{eqnarray}
	p_t &=&\frac {\partial\mathcal{L}}{\partial\dot{t}}=-E \; , \label{pt} \\
	p_\varphi &=&\frac {\partial\mathcal{L}}{\partial\dot{\varphi}}=L \; , \label{pvarphi} \\
	p_r &=&\frac {\partial\mathcal{L}}{\partial\dot{r}} \; , \label{pr} \\
	p_z &=&\frac {\partial\mathcal{L}}{\partial\dot{z}} \; , \label{pz}
	\label{momentos}
\end{eqnarray}
onde $E$ e $L$ são quantidades conservadas \cite{Matos2002}.
Pelas equações (\ref{pt}) e (\ref{pvarphi}), temos
\begin{eqnarray}
	\dot{t} &=& \left( 1- \frac{4J^2}{\mu_s\, r^2} \right) E\, -\, \frac{2J\, L}{\mu_s\, r^2} \; , \label{tponto} \\
	\dot{\varphi} &=& \frac{L}{r^2} + \frac{2J\, E}{\mu_s\, r^2} \; . \label{varphiponto}
\end{eqnarray}
A densidade hamiltoniana dessa partícula é
\begin{eqnarray}
	\mathcal{H} &=& p_\mu \dot{x}^\mu-\mathcal{L}\nonumber \\
	&=& p_t \dot{t}+p_\varphi \dot{\varphi}+p_r \dot{r}+p_z \dot{z}-\mathcal{L} \; .
	\label{Hamiltoniana}
\end{eqnarray}
Consideremos as estrelas das galáxias como partículas teste movendo-se em um movimento circular no plano equatorial ao redor do centro de uma galáxia.
Assim, $\dot{r}=0$ e $\dot{z}=0$.
Além disso, vamos considerar ainda que a densidade hamiltoniana seja normalizada para ser igual a $-\;\frac{1}{2}$ por simplicidade \cite{Matos2002}.\footnote{Observe que qualquer normalização não nula pode ser escolhida.}
Consequentemente, o sistema composto pelas equações
\begin{eqnarray}
	\mathcal{H}+\frac 1 2 &=& 0 \label{Sistema_H_+_partial_H_eq_1} \; ,\\
	\frac \partial {\partial r} \left( \mathcal{H}+\frac 1 2 \right) &=& 0
	\label{Sistema_H_+_partial_H_eq_2}
\end{eqnarray}
permite obter soluções para a energia $E$ e o momento $L$:
\begin{eqnarray}
	E &=& \pm\, 1 \; , \label{solucao_E} \\
	L &=& \mp\, \frac{2J}{\mu_s} \; . \label{solucao_L}
\end{eqnarray}
As soluções (\ref{solucao_E}) e (\ref{solucao_L}) aplicadas nas equações (\ref{tponto}) e (\ref{varphiponto}) fornecem as 4 soluções nulas para a velocidade angular
\begin{eqnarray}
	\Omega&=&\frac {d\varphi}{dt}=\frac{\frac{d\varphi}{d\tau}}{\frac{dt}{d\tau}}=\frac{\dot\varphi}{\dot t} = 0\; .
	\label{velocidade_angular}
\end{eqnarray}
Com a métrica (\ref{eq_3_mazur_BD_phi2mus}) e considerando a prescrição estabelecida por Chandrasekhar $d\tau^2=-ds^2$ \cite{Chandrasekhar1983}, temos
\begin{eqnarray}
	d\tau^2 & = & F_1\; dt^2 \left\{ {1-F_2 \left [ \left( \frac{dr}{dt} \right)^2+ \left( \frac{dz}{dt} \right)^2 \right ]} - F_3 \left( \frac{d\varphi}{dt} - F_4 \right)^2 \right \} \; ,
	\label{dtau2}
\end{eqnarray}
com
\begin{eqnarray}
F_1 &=& \frac{\mu_s^{\, 2}\, r^2}{\mu_s^{\, 2}\, r^2\, -\, 4J^2}\; , \\
F_2 &=& 1-\frac{4J^2}{\mu_s^{\, 2}\, r^2}\; , \\
F_3 &=& \left( r\, -\, \frac{4J^2}{\mu_s^{\, 2}\, r} \right)^2\; , \\
F_4 &=& \frac{2J\mu_s}{\mu_s^{\, 2}\, r^2\, -\, 4J^2} \; .
\end{eqnarray}
Considerando $u^0=\frac{dt}{d\tau}$, a equação (\ref{dtau2}) pode ser reescrita como
\begin{equation}
	1= F_1 \left( u^0 \right)^2 \left( 1-v^2 \right) \; ,
\end{equation}
onde
\begin{equation}
	v^2=v_r^{\; 2}+v_z^{\; 2} + v_\varphi^{\; 2} \; .
\end{equation}
Portanto, a razão entre a velocidade tangencial $v_\varphi$ e a velocidade da luz $c$ pode ser escrita como
\begin{equation}
	\frac{v_{\varphi}}{c}= \frac{2J}{\mu_s\, r} \; .
	\label{velocidade_tangencial}
\end{equation}

\section{Considerações finais}
%e perspectivas futuras}
\label{Considerações}

\begin{figure}[htb]
	\begin{center}
		\includegraphics[width=12cm]{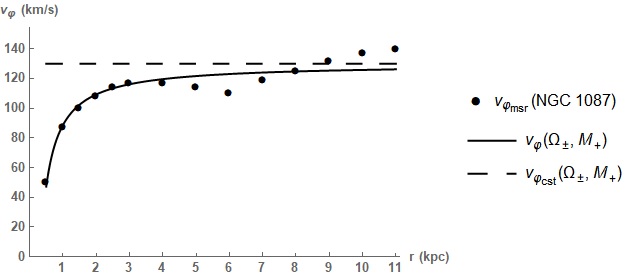}
		\caption{Velocidades das estrelas $v_{\varphi_{msr}}$ da galáxia NGC 1087 \cite{Rubin1980} comparadas com o melhor ajuste da função para a velocidade tangencial $v_{\varphi}$, obtida no trabalho \cite{Santos2017}.
		À medida que $r$ cresce, $v_\varphi$ tende para um valor constante $v_{\varphi_{cst}}$.}
		\label{Grafico_vvarphi-r_NGC_1087}
	\end{center}
\end{figure}

\begin{figure}[htb]
	\begin{center}
		\includegraphics[width=8cm]{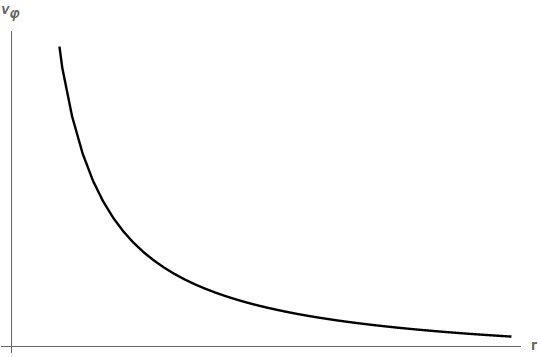}
		\caption{Comportamento da velocidade tangencial $v_\varphi$ das estrelas em função da distância $r$ até o centro de uma galáxia, quando a fonte do espaço-tempo na região dessas estrelas é uma corda cósmica girante na gravitação de Brans-Dicke. Claramente, não há concordância com aquilo que é observado, conforme já mostraram os trabalhos \cite{Zwicky1933, Zwicky1937, Rubin1980}.}
		\label{Grafico_vvarphi-r_Resultado_2023}
	\end{center}
\end{figure}

No trabalho \cite{Santos2017}, foi encontrada uma função para a modelagem da velocidade tangencial das estrelas de algumas galáxias do tipo Sc, considerando que a fonte geradora do espaço-tempo na região onde estão localizadas essas estrelas era simetricamente cilíndrica, como preveem os modelos para as cordas cósmicas girantes.
Agora, o modelo proposto considerou também uma fonte simetricamente cilíndrica, mas com as propriedades características de uma corda cósmica girante, como indica a métrica (\ref{eq_3_mazur_BD_phi2mus}).
Diferente do resultado \cite{Santos2017}, quando $v_{\varphi}\propto r^{b\, >\, 0}$, tendendo a um valor constante, para $r\rightarrow \infty$, e concordando com o que é observado para o comportamento das velocidades das estrelas nas galáxias (veja o gráfico da figura \ref{Grafico_vvarphi-r_NGC_1087}), o resultado obtido aqui é diferente, com $v_{\varphi}\propto \frac{1}{r}$, como mostra a solução (\ref{velocidade_tangencial}), e pode ser visto também no gráfico da figura \ref{Grafico_vvarphi-r_Resultado_2023}.
Portanto, uma fonte simetricamente cilíndrica pode ser um bom modelo para o comportamento das curvas de rotação observadas nas galáxias, mas essa fonte não parece ser uma corda cósmica girante, quando considerado que o espaço-tempo seja aquele previsto pela gravitação de Brans-Dicke.

\begin{comment}
O método apresentado consiste em uma alternativa à resolução do sistema de equações diferenciais acopladas obtido a partir da gravitação de BD, para quando a métrica que define a estrutura a ser estudada é reescrita sob a forma das funções $A_n$.
A Seção \ref{A11} permitiu demonstrar a facilidade e a eficácia no uso da proposta, quando foi aplicada para o estudo das cordas cósmicas girantes com estrutura interna, cujas soluções para o sistema das equações que as definem na gravitação de BD vinham sendo buscadas desde 2018, através dos trabalhos \cite{Chaves2018, Oliveira2021}.

A solução (\ref{solucaophiMazurBD}) aponta para o campo $\phi$ constante, tal como acontece na RG, mas, diferente da gravitação de Einstein, quando $G\phi\sim 10^0$, o resultado encontrado aqui mostra $G\phi\sim 10^{11}$.

No futuro, deve ser explorada uma abordagem considerando também métricas que não mantêm a invariância de Lorentz, ou seja, quando $\beta\neq 2\alpha$.
Para isso, ao invés das soluções (\ref{solucaobeta})-(\ref{solucaophigmunu}), é apropriado o uso das soluções demonstradas no trabalho \cite{Santos2017}.
\end{comment}

\newpage

\bibliographystyle{aip}
\bibliography{referencias_2023-10-08}

%\printbibliography

%\bibliography{referencias_9}

\end{document}